\documentclass[notitlepage,aps,12pt,tightenlines,nofootinbib,showkeys]{revtex4}
\pdfoutput=1

\usepackage{amsmath}
\usepackage{amssymb,amsfonts}
\usepackage{bm}
\usepackage[mathcal]{euscript}
\usepackage{graphicx}
\usepackage{subfigure}
\usepackage{color}

\graphicspath{{figs/}{figs_lo/}}

\newcommand{\mathnotation}[2]{\newcommand{#1}{\ensuremath{#2}}}

\mathnotation{\nr}{n}				
\mathnotation{\npr}{p}				
\mathnotation{\ho}{\varphi}			
\mathnotation{\hoii}{\psi}			
\mathnotation{\surf}{\mathcal{S}}		
\mathnotation{\fol}{\mathcal{F}}		
\mathnotation{\folu}{\fol^{\mathrm{u}}}		
\mathnotation{\fols}{\fol^{\mathrm{s}}}		
\mathnotation{\tmeas}{\mu}			
\mathnotation{\tmeasu}{\tmeas^{\mathrm{u}}}	
\mathnotation{\tmeass}{\tmeas^{\mathrm{s}}}	
\mathnotation{\Nfol}{\#\text{ of singularity data}}
\mathnotation{\parti}{\mathfrak{p}}		
\mathnotation{\bsing}{N_{\mathrm{sep}}}		
\mathnotation{\sing}{N}				
\mathnotation{\per}{i}				

\newcommand{\TN}{Thurston--Nielsen}

\begin{document}

\title{Topology of Chaotic Mixing Patterns}

\author{Jean-Luc Thiffeault}
\email{jeanluc@mailaps.org}
\affiliation{Department of Mathematics, University
  of Wisconsin, Madison, WI 53706, USA}

\author{Matthew D. Finn}
\affiliation{School of Mathematical Sciences, University
  of Adelaide, Adelaide SA 5005, Australia}

\author{Emmanuelle Gouillart}
\affiliation{Unit\'e mixte Saint-Gobain/CNRS \hbox{``Surface du Verre et
  Interfaces,''} 39 quai Lucien Lefranc, 93303 Aubervilliers Cedex,
  France}

\author{Toby Hall}
\affiliation{Department of Mathematical Sciences,
  University of Liverpool, Liverpool L69 7ZL, UK}

\date{\today}


\begin{abstract}

  A stirring device consisting of a periodic motion of rods induces a
  mapping of the fluid domain to itself, which can be regarded as a
  homeomorphism of a punctured surface.  Having the rods undergo a
  topologically-complex motion guarantees at least a minimum amount of
  stretching of material lines, which is important for chaotic mixing.
  We use topological considerations to describe the nature of the
  injection of unmixed material into a central mixing region, which
  takes place at~\emph{injection cusps}.  A topological index formula
  allow us to predict the possible types of unstable foliations that
  can arise for a fixed number of rods.

\end{abstract}

\maketitle

\textbf{
  By stirring a fluid, mixing is greatly enhanced.  By this we mean
  that if our goal is to homogenise the concentration of a substance,
  such as milk in a teacup, then a spoon is an effective way to mix.
  But in many industrial applications, such as food and polymer
  processing, the fluid is very viscous, so that stirring is difficult
  and costly.  Hence, insight into the types of stirring that lead to
  good mixing is valuable.  We explain how the mixing pattern --- the
  characteristic shape traced out by a blob of dye after a few
  stirring periods --- is tightly connected with topological
  properties of the stirring motion.  In particular, we can enumerate
  the allowable number of pathways where material gets injected into the
  mixing region, as a function of the number of stirring rods.
}

\section{Stirring with Rods}
\label{sec:stirring}

\begin{figure}
  \begin{center}
    \includegraphics[width=.5\textwidth]{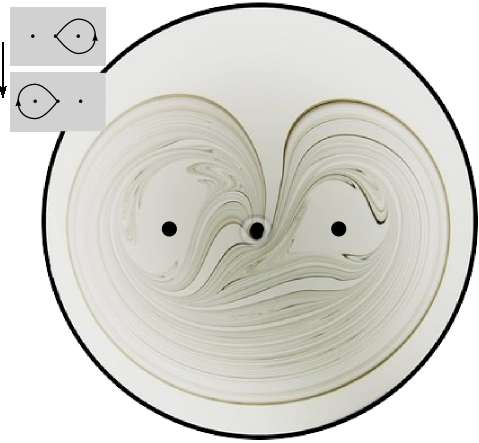}
  \end{center}
  \caption{The figure-eight stirring protocol.  The inset shows the
    sequence of rod motions.  The central black circle is the stirring
    rod, and the other two circles locate the position of regular
    islands, which serve as `ghost rods.'  (Experiments by E.
    Gouillart and O. Dauchot, CEA Saclay~\cite{Gouillart2007}.)}
  \label{fig:fig8_protocol}
\end{figure}

A \emph{rod stirring device}, where a number of rods are moved around
in a fluid, is the most natural and intuitive method of stirring.  The
number of rods, their shape, and the nature of their motion constitute a
\emph{stirring protocol}.  For example, Fig.~\ref{fig:fig8_protocol}
shows the result of stirring with the \emph{figure-eight} protocol,
whereby a single rod in a closed vessel traces a lemniscate shape.
The Reynolds number is very small, so that the fluid (sugar syrup) is
in the Stokes regime, where inertial forces are negligible and
pressure and viscous forces are in balance.  Because of the shape
of the rod and container, three-dimensional effects are negligible.
The fluid is the pale background, and a blob of black ink has been
stretched by a few periods of the rod motion.
The evident filamentation of the blob is characteristic of
\emph{chaotic advection}~\cite{Aref1984,Aref2002}, which
greatly enhances mixing effectiveness in viscous flows.

\begin{figure}
\begin{center}
\subfigure[]{
  \includegraphics[width=.35\textwidth]{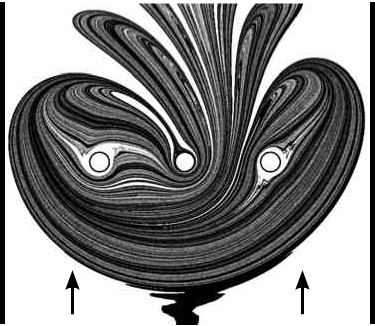}
\label{fig:openBF}
}\hspace{2em}%
\subfigure[]{
  \includegraphics[width=.35\textwidth]{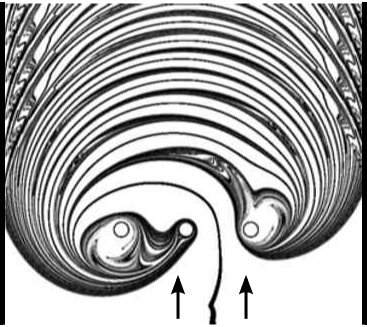}
\label{fig:openBS}
}
\end{center}
\caption{Numerical simulations of flow in a channel with a figure-eight
  rod stirring protocol. (a) Injection cusp against flow; (b)
  Injection cusp facing flow.  The flow direction is from the bottom
  to the top of the figures.}
\label{fig:channel_exp}
\end{figure}

We aim to understand the features of rod stirring protocols such as
those depicted in Figs.~\ref{fig:fig8_protocol},
\ref{fig:channel_exp}, and~\ref{fig:s1s2s-3} from topological
considerations.  The new topological features that we discuss can be
divided in two broad categories: (i) The injection cusps and their
role; (ii) The identification of higher-pronged singularities directly
in flow simulations.  Though both of
these concepts are familiar from the topological study of surface
homeomorphisms~\cite{Fathi1979,Thurston1988}, we interpret them here
in light of practical stirring protocols.  Our study is a natural
continuation of the original investigation by \citet{Boyland2000} and
the subsequent work of \citet{Gouillart2007}.  Our goal is to refine
the previous approaches by looking for more detailed features of the
\TN\ classification in real fluid flows.

As an illustration, we examine the topological features visible in
Fig.~\ref{fig:fig8_protocol}.  The two unmixed regions in the centre
of the loops of the figure-eight play the role of two extra rods,
called \emph{ghost rods}~\cite{Gouillart2006}, so that we can regard
this protocol as effectively involving three rods.  Three rods is the
minimum needed to guarantee exponential stretching of material
lines~\cite{Boyland2000,Boyland2003}, and for the figure-eight
protocol the length of material lines grows by at least a factor
of~$(1+\sqrt{2})^2$ at each period.  This is the first feature
that can be understood from the topology of the rod motion: it places
a lower bound on the \emph{topological entropy}, which is closely
related to the rate of stretching of material lines in two dimensions.
This aspect has been well
studied~\cite{Boyland2000,Boyland2003,Vikhansky2003,%
  MattFinn2003,MattFinn2003b,Thiffeault2005,Gouillart2006,%
  Thiffeault2006,Kobayashi2007,Binder2008,MattFinn2006,%
  MattFinn2007}.

Less well studied is another crucial feature obvious in
Fig.~\ref{fig:fig8_protocol}: the unmixed fluid (white) is injected
into the mixing region (kidney-shaped darker region) from the top part
of the region, where a cusp is clearly visible.  We say that the
figure-eight protocol has one \emph{injection cusp}.  In fact, for
three rods the situation in Fig.~\ref{fig:fig8_protocol} is typical;
for instance, a kidney-shaped mixing region is also evident in the
efficient stirring protocol of \citet{Boyland2000}.  The nature of the
injection of unmixed material into the central region has a profound
impact on mixing rates, as was shown in recent
experiments~\cite{Gouillart2007}.  In these experiments the rate of
injection of unmixed material into the central region of a stirring
device dramatically limited the efficiency of mixing.  This injection
took place along injection cusps as described here, hence the number
of injection cusps and their positions are clearly important for
mixing.

The importance of injection cusps is even more apparent when dealing
with \emph{open flows}.  Open flows, as opposed to flows in closed
vessels, involve fluid that enters and subsequently exits a mixing
region.  This situtation is very common in industrial settings, since
it allows continuous operation without the need to empty the vessel.
Typical fluid particles only remain in the mixing region for a short
time.  Figure~\ref{fig:channel_exp} shows the figure-eight stirring
protocols in an
open channel.  The only difference between cases~(a) and~(b) is the
direction of rotation of the rods.  In a closed vessel, reversing the
direction of rotation merely moves the injection cusp from top to
bottom, but in an open flow it moves the injection cusp either facing
the flow or against it.  In Fig.~\ref{fig:openBF}, the injection cusp
is downstream, so the passive scalar enters the mixing region from
behind.  This is clear from the dye filament at the bottom, which
skirts the mixing region from the right.  In contrast,
Fig.~\ref{fig:openBS} shows the opposite case, where the injection
cusp is upstream.  In that case the dye is drawn directly into the
mixing region.  Clearly these two cases have radically different
mixing properties, as evidenced by the different dye patterns
downstream.  Thus, being able to predict the number and nature of
injection cusps from the rod motion is of crucial importance.  The
qualitative details of Fig.~\ref{fig:channel_exp} are independent of
the manner in which dye is injected.  The extra injection cusps
visible downstream in Fig.~\ref{fig:openBF} are an artifact of the
open-flow configuration: they are images of the single injection cusp
being advected downstream by the mean flow at each period.  The
results presented in the rest of the paper apply rigorously only to
closed flows, but it is evident from Fig.~\ref{fig:channel_exp} that
many qualitative features carry over from closed to open flows.

In this paper we will see that the number and position of injection
cusps depends on the topology of the rod motion.  The number of
injection cusps will depend on the number of rods (or ghost rods), but
for a fixed number of rods only a few configurations are possible.
This is because the number of injection cusps is constrained by a
topological \emph{index formula}.  \citet{Jana1994b} studied the
impact of the topology of streamlines on chaotic advection.  They used
the Euler--Poincare--Hopf formula (see Section~\ref{sec:index}) to
determine the allowable fixed-point structure of steady velocity
fields.  When the flow is time-dependent, instantaneous fixed points of
the velocity field mean little for chaotic advection (except when the
time dependence is weak).  Our study addresses arbitrary time-periodic
flows by studying the mapping of fluid elements (the Lagrangian map)
directly, from a topological perpective.

The paper is organised as follows.  In Section~\ref{sec:TN} we present
some necessary mathematical background, in particular the idea of
\emph{pseudo-Anosov} stirring protocols.  We identify mathematical
objects such as the \emph{unstable foliation} in a specific
fluid-dynamical example.  The unstable foliation associated with a
pseudo-Anosov protocol is the central object of our study.
In Section~\ref{sec:index} we use
a topological index formula to enumerate the possible unstable
foliations for a given number of stirring rods, and
Section~\ref{sec:trm} is devoted to a stirring device that exhibits
a \emph{hyperbolic} injection cusp, that is a cusp associated with the
unstable manifold of a hyperbolic orbit.  Such a cusp is less obviously
identifiable from flow visualization, but can still be inferred by
examining the topological properties of the stirring protocol.
Finally, we summarise and discuss our work
in Section~\ref{sec:discussion}.

\section{Pseudo-{A}nosov Stirring Protocols}
\label{sec:TN}

In this section we will translate the physical system --- a stirring
protocol --- into objects suitable for mathematical study.  As mentioned
in Section~\ref{sec:stirring}, our focus will be on very viscous
flows, where stirring and mixing is challenging, and we consider the
flow to be essentially two-dimensional, as is typical of shallow or
stratified flows.  In these circumstances, a periodic motion of rods
leads to a periodic velocity field.  (Note that the rods may end up
permuted amongst themselves at the end of each period.)  This velocity
field will induce a motion of fluid elements from their position at
the beginning of a period to a new position at the end.  Crucially,
the fluid elements do not necessarily return to the same position --- if
they did, this would be a very poor stirring protocol indeed!

A periodic stirring protocol in a two-dimensional flow thus induces a
\emph{homeomorphism}~$\ho$ from a surface~\surf\ to itself.  A
homeomorphism is an invertible continuous map whose inverse is also
continuous.  In our case,~$\ho$ describes the mapping of fluid
elements after one full period of stirring, obtained from solving the
Stokes equation, and~$\surf$ is the \emph{disc} with holes (or
\emph{punctures}) in it, corresponding to rods.  We treat rods as
infinitesimal punctures (see Fig.~\ref{fig:rodsing_point}), since
topologically this makes no difference.  As a special case, rods that
remain fixed are often called \emph{baffles}.  Topologically speaking,
moving rods and fixed baffles are the same: they are holes in the
surface~$\surf$.  However, the homeomorphism~$\ho$ acts on them
differently: the moving rods can be permuted, whilst the fixed baffles
remain in place.  Here, we shall not make a distinction between
stirring rods and fixed baffles, and refer to both as stirring rods.
Hence, a few stirring rods may be fixed by some stirring protocols,
such as the figure-eight protocol in Fig.~\ref{fig:fig8_protocol}
which fixes two rods (the islands, or ghost rods).  The outer boundary
of the disc is invariant under~$\ho$, corresponding to the no-slip
boundary condition.

Our task is to categorise all possible~$\ho$ that lead to good mixing.
This requires defining both what we mean by `categorise' and `good
mixing.'  The categorisation will be done up to isotopy, which is a
way of defining the topological equivalence of homeomorphisms.  Two
homeomorphisms~$\ho$ and~$\hoii$ are \emph{isotopic} if~$\hoii$ can be
continuously `reached' from $\ho$ without moving the rods.  If we
imagine the two-dimensional fluid as a rubber sheet, this means that
the two configurations attained by the sheet after application of
either~$\ho$ or~$\hoii$ are the same, up to deformation of the sheet.
In that case, we write~$\ho \simeq \hoii$.  Obviously, topology is
unconcerned by hydrodynamic details and only deals with coarse
properties of the homeomorphisms.

To pursue the categorisation, we invoke the \TN\ (TN) classification
theorem~\cite{Fathi1979,Thurston1988}, which describes the range of
possible behaviour of a homeomorphism~$\ho$.  Specifically, the
theorem says that $\ho$ is isotopic to a homeomorphism~$\ho'$,
where~$\ho'$ is either \emph{finite-order}, \emph{reducible}, or
\emph{pseudo-Anosov}.  These three cases give the \emph{isotopy class}
of~$\ho$, and~$\ho'$ is called the TN representative of the isotopy
class.  Finite-order means that~$\ho'$ is periodic (that is,
$\ho'^m=\text{identity}$ for some integer~$m>0$), and this cannot give
good mixing since it implies nearby fluid elements will periodically
come back near each other.  The reducible case implies there are
regions of fluid that remain invariant under~$\ho'$, which again is
terrible for mixing since there are then regions that do not mix with
each other.  For both finite-order and reducible~$\ho'$, the actual
homeomorphism~$\ho$ could in practice exhibit much more complicated
properties than~$\ho'$, but this cannot be inferred by the rod motion
itself.  In the viscous flows that we have studied, $\ho$ and $\ho'$
appear to have similar properties.

The third case, when $\ho'$ is pseudo-Anosov (or pA), is both the most
interesting mathematically and most relevant for mixing.  In fact, we
will define `good mixing' as $\ho'$ having the pA property.
Mathematically, a pA homeomorphism~$\ho'$ leaves invariant a
transverse pair of measured singular foliations, $(\folu,\tmeasu)$ and
$(\fols,\tmeass)$, such
that~$\ho'(\folu,\tmeasu)=(\folu,\lambda\,\tmeasu)$
and~$\ho'(\fols,\tmeass)=(\fols,\lambda^{-1}\tmeass)$, for
\emph{dilatation} $\lambda>1$.  (The logarithm of the dilatation is
the \emph{topological entropy}.)  Here~$\tmeasu$ and~$\tmeass$ are the
transverse measures for their respective foliation.  There are several
terms that need explaining in this definition, and we will illustrate
what they mean by an example.

\begin{figure}
\begin{center}
  \includegraphics[width=.45\textwidth]{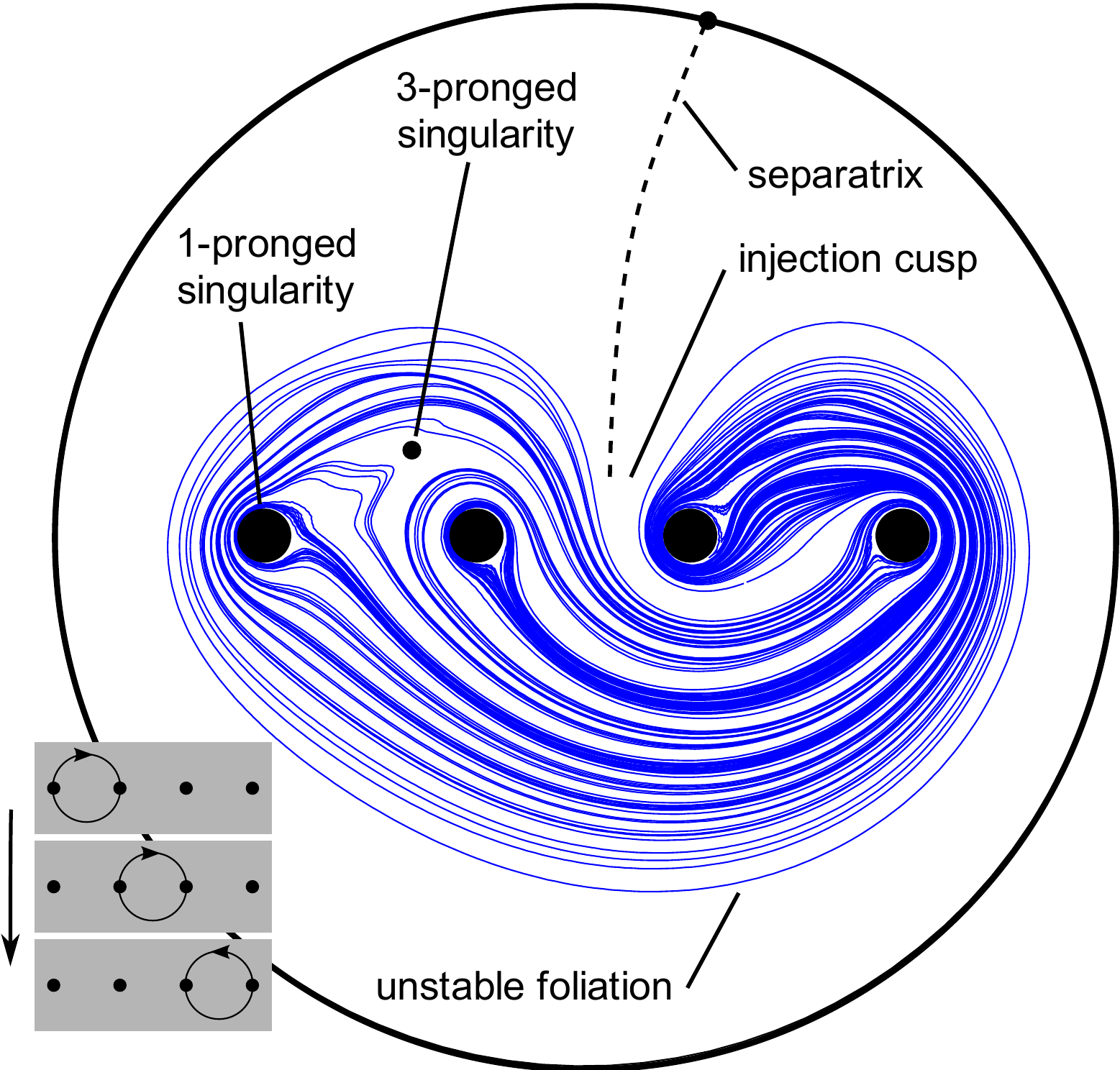}
\end{center}
\caption{Numerical simulation of the four-rod periodic stirring
  protocol~$\sigma_1\sigma_2\sigma_3^{-1}$ in a viscous flow,
  corresponding to three rod interchanges for each period (inset).
  This sequence of interchanges forces the flow to be isotopic to a
  pseudo-Anosov homeomorphism.  A material line advected for six
  periods reveals leaves of the unstable foliation~$\folu$.  The
  foliation exhibits four 1-pronged singularities around the rods, one
  interior 3-pronged singularity (in the region indicated by a dot,
  between and just above the first and second rods), and a separatrix
  attached to the disc's outer boundary, visible as an injection cusp
  along from the top into the mixing region.  (The position of the
  singularities and separatrix are approximate.)  Compare with
  Fig.~\ref{fig:sing_us}, which shows
  typical leaves of the foliation in the neighbourhood of
  singularities.}
\label{fig:s1s2s-3}
\end{figure}

\begin{figure}
\begin{center}
\subfigure[]{
  \centering\includegraphics[width=.28\textwidth]{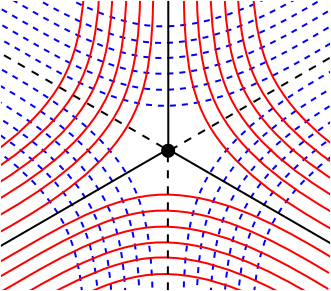}
  \label{fig:3prongsing_us}
}\hspace{2em}
\subfigure[]{
  \centering\includegraphics[width=.28\textwidth]{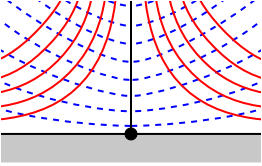}
  \label{fig:boundsing_us}
}\hspace{2em}
\subfigure[]{
  \centering\includegraphics[width=.28\textwidth]{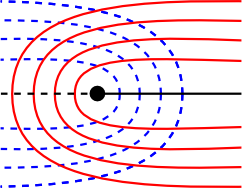}
  \label{fig:rodsing_point}
}
\end{center}
\caption{For the unstable foliation~$\folu$ and stable
  foliation~$\fols$: (a) The neighbourhood of a 3-pronged singularity;
  typical leaves of~$\folu$ are shown as solid lines, dashed lines for
  those of~$\fols$.  (b) The neighbourhood of a boundary singularity.
  A separatrix of~$\folu$ emanates from the singularity, and
  separatrices for~$\folu$ and~$\fols$ alternate around the boundary.
  (c) We regard a rod as an infinitesimal point (or puncture),
  corresponding to a 1-pronged singularity.  The leaves of the
  foliations fold and meet at separatrices attached to the rod.}
\label{fig:sing_us}
\end{figure}

Figure~\ref{fig:s1s2s-3} shows the result of a numerical simulation
of a two-dimensional viscous (Stokes) flow.  The container is
circular, and the fluid is stirred with four rods, shown aligned
horizontally in the centre.  (The velocity field for these simulations
was determined using a fast spectrally-accurate complex variable
method~\cite{MattFinn2003b}, and the particle advection computed with
a high-order Runge--Kutta scheme.)  The inset illustrates the motion
of the stirring rods: they are successively interchanged with their
neighbour, in the direction shown.  This protocol is
written~$\sigma_1\sigma_2\sigma_3^{-1}$, where~$\sigma_i$ denotes the
clockwise interchange of rod~$i$ with rod~$(i+1)$, and~$\sigma_i^{-1}$
its anticlockwise
counterpart~\cite{Birman1975,Boyland2000,Thiffeault2005}.  Thus,
$\sigma_i^{-1}$ is the inverse operation to~$\sigma_i$.  The
subscript~$i$ in~$\sigma_i^{\pm 1}$ refers to the physical position of
a rod from left to right, and does not label a specific rod.  The
collection~$\sigma_i^{\pm 1}$, $i=1,\dots,\nr-1$, generates the
\emph{braid group} on~$\nr$ strands.  We apply the~$\sigma_i^{\pm 1}$
operations, called \emph{braid group generators}, in temporal order
from left to right.  Repeated generators are written as powers, as in
$\sigma_i\sigma_i = \sigma_i^2$.  This sequence of generators, called
a \emph{braid word}, defines our stirring protocol, which gives us the
homeomorphism~$\ho$ after we solve the Stokes equations for a viscous
fluid.  Table~\ref{tab:protocols} defines the protocols used in this
paper in terms of braid group generators.
\begin{table}
  \caption{The pseudo-Anosov stirring protocols used in this paper,
    defined in terms of braid group generators.  The generator
    $\sigma_i$ denotes the clockwise interchange of rod~$i$ with
    rod~$(i+1)$, and~$\sigma_i^{-1}$ its
    anticlockwise counterpart, where~$i$ indicates the physical
    position of a rod in the sequence from left to right. The
    generators are read from left to right in time.}
\label{tab:protocols}
\medskip
\begin{ruledtabular}
\begin{tabular}{cccl}
punctures & protocol & injection cusps & figure \\[2pt]
\hline
3 & $\sigma_2^{-2}\,\sigma_1^2$ & 1
& Fig.~\ref{fig:fig8_protocol}, the figure-eight protocol \\
4 & $\sigma_1\sigma_2\sigma_3^{-1}$ & 1
& Fig.~\ref{fig:s1s2s-3} \\
4 & $\sigma_1\sigma_2^{-1}\sigma_3\sigma_2^{-1}$ & 2
& Fig.~\ref{fig:s1s-2s3s-2} \\
5 & $\sigma_1\sigma_2^{-1}\sigma_3\sigma_4^{-1}\sigma_3\sigma_2^{-1}$
& 3 & Fig.~\ref{fig:s1s-2s3s-4s3s-2} \\
5 & $\sigma_1\sigma_2\sigma_3\sigma_2^3\sigma_3^2
\sigma_4\sigma_2\sigma_3\sigma_2^3$ & 1 & Fig.~\ref{fig:trm}
\end{tabular}
\end{ruledtabular}
\end{table}

The homeomorphism~$\ho$, by the \TN\ theorem, is isotopic to the TN
representative~$\ho'$, which is pseudo-Anosov in this case.
Remarkably, in viscous flows there is often very little visual
difference between the action of~$\ho$ and~$\ho'$, and we find many
features of~$\ho'$ reflected directly in Fig.~\ref{fig:s1s2s-3}.  This
need not be the case in general: the dynamics of~$\ho$ is at least as
complicated as that of~$\ho'$, in a precise
sense~\cite{Handel1985,Boyland1999,Boyland2000}, but can in practice
be considerably more so.

The folded lines in the background of Fig.~\ref{fig:s1s2s-3} are a
small material closed loop that was evolved for six full periods of
the stirring protocol.  As the material line is evolved for more and
more periods, it converges to the unstable foliation, tracing
out~$\folu$ and allowing us to visualise it as a striated pattern.
(The stable foliation~$\fols$ is invisible in such
experiments, so we shall not have much use for it here.)  That the
foliation~$\folu$ remains invariant under~$\ho'$ means that at each
application of~$\ho'$ the bundle of lines in Fig.~\ref{fig:s1s2s-3} is
unchanged, except for the lines getting denser: each application
multiplies the number of lines in a given region by~$\lambda$, the
dilatation. (A local count of the line density is what the invariant
measure~$\tmeasu$ gives us.)  The unstable foliation~$\folu$ is thus
the object that captures the essence of stretching and folding in a
chaotic flow.

Finally, the `pseudo' in pseudo-Anosov is tied to the `singular' in
singular foliation.  An Anosov homeomorphism, such as Arnold's cat
map~\cite{ArnoldAvez}, can only exist on surfaces of zero Euler
characteristic such as the torus.  This is because the torus is a
surface on which a nonvanishing vector field (in this case the
foliation) can be smoothly combed.  For other surfaces, such as our
punctured disc, the best one can do is to comb the foliation and leave
some singularities.
Three such singularities are shown in Fig.~\ref{fig:sing_us}.  The
singularities are characterised by the number of \emph{prongs}
associated with them.
The prongs are separatrices, emanating from the singular point, around
which the foliation branches.  For instance,
Fig.~\ref{fig:3prongsing_us} shows a 3-pronged singularity: the leaves
of the unstable foliation~$\folu$ (solid lines) branch around the
singularity in a triangular pattern around three separatrices.  The
dashed lines show leaves of the stable foliation,~$\fols$.
Figure~\ref{fig:boundsing_us} shows a separatrix
of~$\folu$ attached to the outer boundary of the disc at a singularity.
Unstable and stable separatrices alternate around the boundary.
Figure~\ref{fig:rodsing_point}
shows a 1-pronged singularity, which occurs around rods, as we will
see below.

The stirring device in Fig.~\ref{fig:s1s2s-3} shows a total
of six singularities in~$\folu$.  Most obviously, there is a 1-pronged
singularity around each of the four rods, because the leaves are
folded around each rod and meet at a separatrix, as in
Fig.~\ref{fig:rodsing_point}.  Next, there is a 3-pronged singularity,
in the region marked with a dot in the picture (this dot is not a
rod).  That the singularity has three separatrices is evident if one
tries to extend 
the foliation near the singularity: three bundles of leaves will meet
at a point, which must then be a singularity.  Finally, there is a
separatrix connected to the disc's outer boundary at a singularity, as
in Fig.~\ref{fig:boundsing_us}, though this is not as easy to see.
From Fig.~\ref{fig:boundsing_us} we expect that a boundary singularity
will be manifested as a cusp in~$\folu$, a consequence of the
separatrix emanating from the boundary, and leading to injection of
material into the mixing region (Fig.~\ref{fig:s1s2s-3}).

In the pA case, almost every aspect of the \TN\ classification theorem
can thus be identified directly in Fig.~\ref{fig:s1s2s-3}.  Rods and
the outer boundary possess singularities, and often other (interior)
singularities arise in the flow itself, such as the 3-pronged
singularity indicated by a dot in Fig.~\ref{fig:s1s2s-3}.  We will
refer to singularities not associated with rods or the outer boundary
as \emph{interior singularities}.  Our reference protocol of
Fig.~\ref{fig:s1s2s-3} demonstrates that the unstable
foliation~$\folu$ and its singularities embody many
important features of the underlying flow.  We shall thus take the
unstable foliation as the central focus of our study.

\section{Singularities of the Foliation}
\label{sec:index}

An unstable foliation must satisfy three rules if it is to support a
stirring protocol corresponding to a pseudo-Anosov
homeomorphisms~\cite{PennerHarer}:
\begin{enumerate}
\item Every stirring rod (or ghost rod, such as the islands in the
  figure-eight protocol) must be enclosed in a 1-pronged singularity
  (Fig.~\ref{fig:rodsing_point}).  This is a physical requirement:
  1-pronged singularities are the mathematical consequence of physical
  stirring, since the unstable foliation wraps around the rod.  A rod
  enclosed in a higher-pronged singularity makes the rod irrelevant to
  stirring, since the singularity would exist regardless of the
  rod. Hence, we disregard this possibility.
  \label{item:1}
\item The outer boundary of the disc contains at least one separatrix
  of~$\folu$, as in
  Fig.~\ref{fig:boundsing_us}.  The number of separatrices on the
  outer boundary corresponds to the number of injection cusps into the
  mixing region.
\item The smallest number of prongs an interior singularity can have is
  3. This is because 2-pronged singularities are just regular
  points (they are not `true' singularities of the foliation), and
  pseudo-Anosovs do not have 1-pronged singularities away from
  punctures and boundary components.
  \label{item:3}
\end{enumerate}

We will use these three rules to limit the number of allowable
\emph{singularity data} of the unstable foliation~$\folu$.
The singularity data of a foliation~$\fol$ is the sequence
$(\bsing,\sing_3, \sing_4, \ldots)$, where $\bsing$ is the number of
separatrices on the outer boundary, and $\sing_\npr$ is the number of
interior $\npr$-pronged singularities for each $\npr\ge 3$.  To
enumerate the possible distinct singularity data of the unstable
foliation for $\nr$ rods, we use a standard \emph{index formula},
which relates the nature of the singularities to a topological
invariant, the \emph{Euler characteristic} of the disc,
\emph{$\chi_{\mathrm{disc}}=1$}.%
\footnote{More generally, the Euler characteristic of a surface of
  genus~$g$ with~$b$ boundaries is~$2-2g-b$, where the genus is the
  number of `handles' attached to a sphere.  In
  particular,~$\chi_{\mathrm{sphere}}=2$,~$\chi_{\mathrm{disc}}=2-1=1$,
  and~$\chi_{\mathrm{torus}}=2-2\cdot 1=0$.}
This index formula says that
\begin{equation}
  \nr - \bsing - \sum_{\npr\ge3}(\npr - 2)\sing_\npr =
  2\chi_{\mathrm{disc}} = 2.
  \label{eq:EP}
\end{equation}
Formula~\eqref{eq:EP} is a well-known extension to foliations of the
classical Euler--Poincar\'e--Hopf formula for vector
fields~\cite{Milnor,Thurston3D1} (see for example page 1352
of~\cite[p.~1352]{BandBoyland2007}).  It is important to note that the
only positive contribution to the left hand side of~\eqref{eq:EP} is
the term~$\nr$, the number of rods. Hence if there is a large number
of rods, there must either be a large number of singularities and
boundary separatrices, or a small number of singularities some of
which have many prongs.

In order for~\eqref{eq:EP} to be satisfied, the boundary separatrices
and interior singularities must contribute~$(2 - \nr)$ to the left
hand side. For~$\nr=3$, the only possibility is to have a single
separatrix on the boundary. There is therefore unique singularity data
for three stirring rods in a disc, corresponding to the kidney-shaped
region in Fig.~\ref{fig:fig8_protocol}: the separatrix on the boundary
corresponds to a single injection cusp into the mixing region.

\begin{figure}
  \begin{center}
    \subfigure[]{
      \includegraphics[width=.45\textwidth]{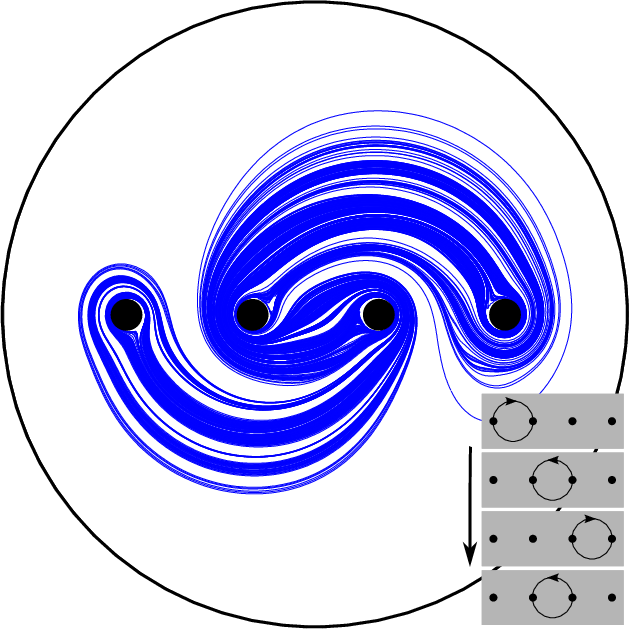}
      \label{fig:s1s-2s3s-2}
    }\hspace*{2em}
    \subfigure[]{
      \includegraphics[width=.45\textwidth]{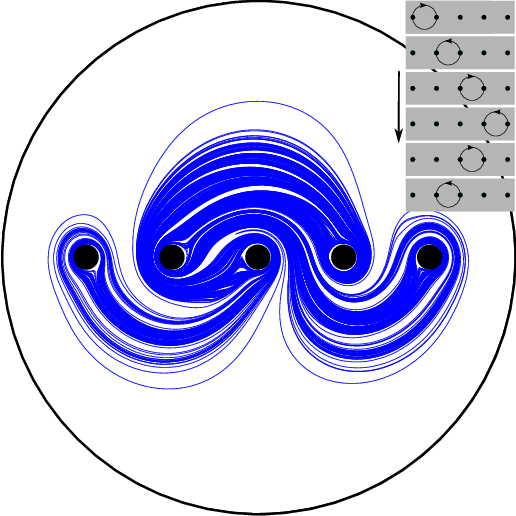}
      \label{fig:s1s-2s3s-4s3s-2}
    }
  \end{center}
  \caption{Numerical simulations of (a) the stirring
    protocol~$\sigma_1\sigma_2^{-1}\sigma_3\sigma_2^{-1}$ for 4 rods,
    with two injection cusps; (b) the stirring
    protocol~$\sigma_1\sigma_2^{-1}\sigma_3\sigma_4^{-1}\sigma_3\sigma_2^{-1}$
    for 5 rods, with three injection cusps.}
  \label{fig:morerods}
\end{figure}

For~$\nr=4$, the boundary separatrices and interior singularities
must contribute~$-2$ to the left hand side of~\eqref{eq:EP}, which
means that there must either be two separatrices on the boundary, or
one separatrix on the boundary and an interior 3-pronged
singularity. These two cases correspond to the unstable foliations of
the second and third protocols in Table~\ref{tab:protocols}, as
depicted in Figs.~\ref{fig:s1s-2s3s-2} and~\ref{fig:s1s2s-3}.

As we add more rods, there are more possibilities for the singularity
data. For example, Fig.~\ref{fig:s1s-2s3s-4s3s-2} shows a protocol for
five rods ($\nr=5$) with three boundary separatrices ($\bsing=3$),
corresponding to three injection cusps, and no interior singularities.

The maximum number of interior singularities occurs when there is a
single boundary separatrix. The rods and this separatrix then
contribute~$\nr-1$ to the left hand side of~\eqref{eq:EP}, and
so~\eqref{eq:EP} can be satisfied if there are~$\nr-3$ interior
3-pronged singularities. That is, the maximum possible number of
interior singularities occurs for~$\sing_3=\nr-3$. On the other hand, in
order to have no interior singularities it is necessary to
have~$\nr-2$ boundary separatrices. This corresponds to the maximum
possible number of injection cusps into the stirring region.

Table~\ref{tab:tttypes} provides a complete summary of the allowable
singularity data for the first few values of~$\nr$.
\begin{table}
  \caption{The allowable singularity data $(\bsing,\sing_3, \sing_4,
    \ldots)$ for~$\nr$ rods.  Each
    rod has a 1-pronged singularity, and formula~\eqref{eq:EP} must be
    satisfied.  $\bsing$ gives the number of separatrices on the outer
    boundary, and $\sing_\npr$ gives the number of $\npr$-pronged
    interior singularities.}
\label{tab:tttypes}
\medskip
\begin{ruledtabular}
\begin{tabular}{lcccc}
$\nr$ & $\bsing$ & $\sing_3$ & $\sing_4$ & $\sing_5$ \\[2pt]
\hline
3 & 1 &  &  &  \\
\hline
4 & 2 & 0 &  &  \\
4 & 1 & 1 &  &  \\
\hline
5 & 3 & 0 & 0 &  \\
5 & 2 & 1 & 0 &  \\
5 & 1 & 2 & 0 &  \\
5 & 1 & 0 & 1 &  \\
\hline
6 & 4 & 0 & 0 & 0 \\
6 & 3 & 1 & 0 & 0 \\
6 & 2 & 2 & 0 & 0 \\
6 & 2 & 0 & 1 & 0 \\
6 & 1 & 3 & 0 & 0 \\
6 & 1 & 1 & 1 & 0 \\
6 & 1 & 0 & 0 & 1
\end{tabular}
\end{ruledtabular}
\end{table}%
%
%
\begin{figure}
  \begin{center}
      \includegraphics[width=.6\textwidth]{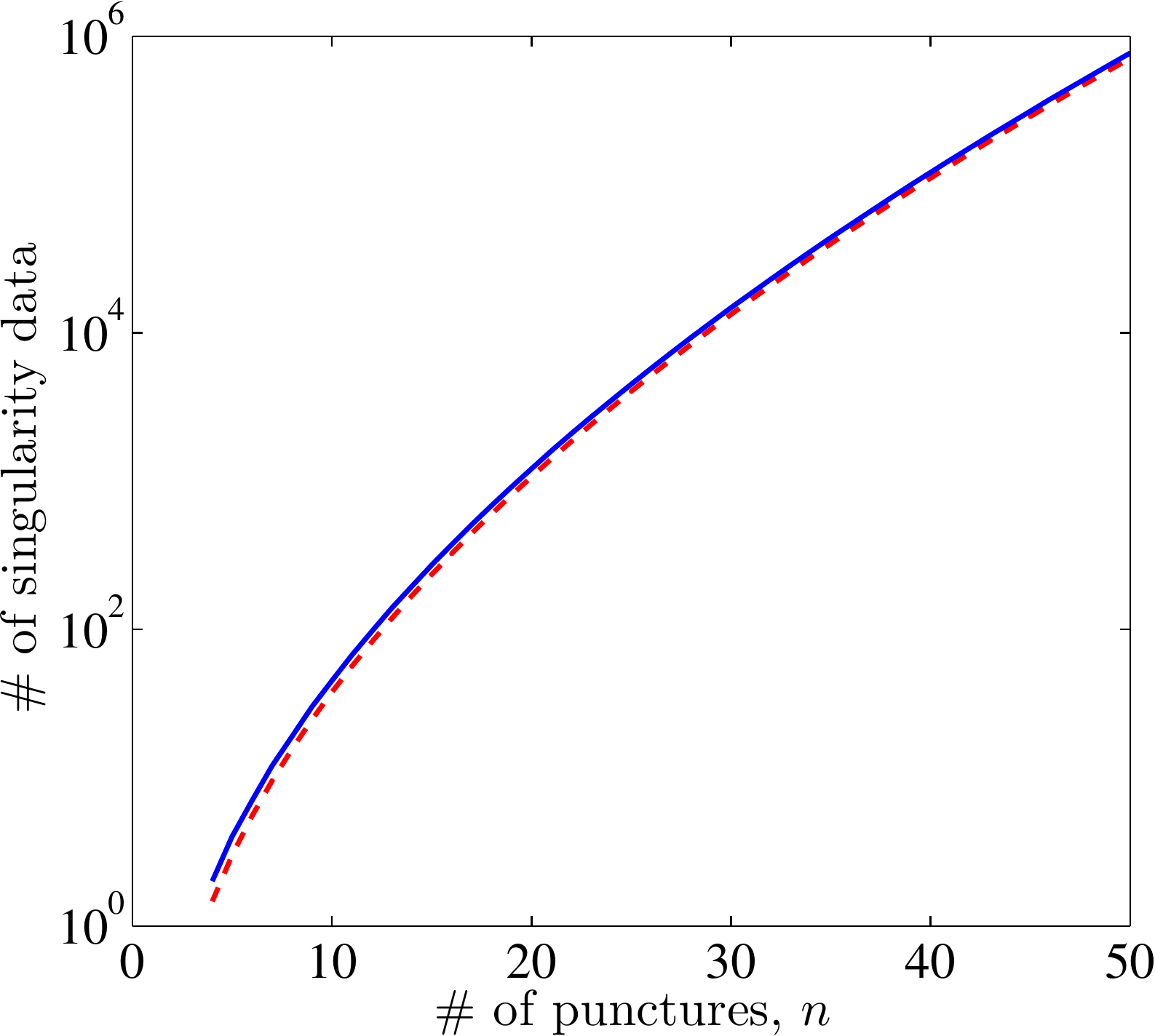}
  \end{center}
  \caption{The number of distinct singularity data increases rapidly
    with the number of punctures (or rods).  The solid line is exact
    and is obtained by summing partition functions as in~\eqref{eq:Nfol};
    the dashed line is the asymptotic form~\eqref{eq:Nfola}.}
  \label{fig:Nfol}
\end{figure}%
The number of possible distinct singularity data for a foliation
increases sharply with the number of rods~$\nr$, as is evident in
Fig.~\ref{fig:Nfol} (solid line).  A convenient expression for the
number of singularity data is
\begin{equation}
  \Nfol = \sum_{k=0}^{\nr-3} \parti(k)
  \label{eq:Nfol}
\end{equation}
where~$\parti(k)$ is a \emph{partition
  function}~\cite{AndrewsPartitions}.  The partition function counts
how many distinct ways positive integers can sum to~$k$,
with~$\parti(0)$ defined as~$1$:
\begin{equation}
  \parti(k) = \text{\# of elements in the set }
    \Bigl\{S\subseteq{\mathbb{Z}}^+\,:\,\sum_{i \in S}i = k\Bigr\}.
\end{equation}
The partition function has no simple exact closed form.  To find the
asymptotic form of~\eqref{eq:Nfol} for large~$\nr$, we can use the
Hardy--Ramanujan asymptotic form for~$\parti(k)$, and replace the sum
by an integral, to get
\begin{equation}
  \Nfol \sim \frac{1}{2\pi \sqrt{2\, (n-3)}}
  \exp\left(\pi\sqrt{2(n-3)/3}\right),\qquad \nr \gg 1.
  \label{eq:Nfola}
\end{equation}
The dashed line in Fig.~\ref{fig:Nfol} shows that the aymptotic form
captures the correct order of magnitude for large~$\nr$.

\section{Hyperbolic Injection Cusps}
\label{sec:trm}

Injection cusps are not always as plainly visible as in the cases
presented thus far.  However, as we will see in this section, their
presence can still be inferred by examining the topological properties
of the rod motion, including if necessary the motion of ghost rods.
This section also helps to clarify the type of topological information
that can be gleaned from the motion of
rods~\cite{Gouillart2006,Stremler2007,Binder2008}:
\begin{itemize}
\item If the motion of the rods themselves forms a pseudo-Anosov
  braid, then the rod motion yields interesting topological
  information even with no knowledge of ghost rods.  This is the case
  with all the protocols in the paper thus far (except the
  figure-eight, which involves two obvious ghost rods).  The
  topological information is then ``robust,'' in the sense that it
  does not depend on the specific hydrodynamics of the fluid.
\item If the motion of the rods does not imply a pseudo-Anosov, such as when
  there is only one or two rods~\cite{Boyland2000}, and chaotic
  behaviour is observed regardless, then ghost rods must be included.
  This means either looking for regular islands (as in the example in
  this section) or for unstable periodic orbits.  However, the
  presence and location of such periodic structures depend on the
  specific hydrodynamic model (here Stokes flow for a viscous fluid).
\end{itemize}

\begin{figure}
\subfigure[]{
  \includegraphics[width=.4\textwidth]{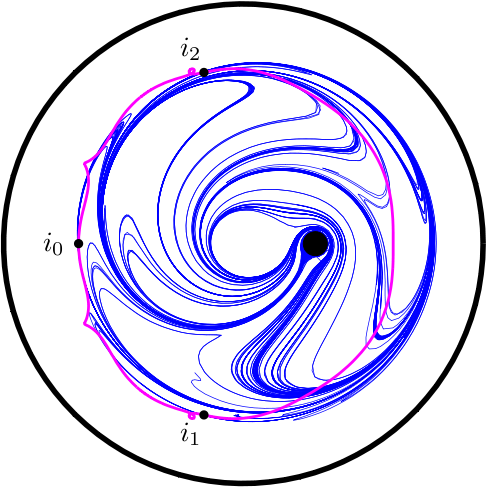}
  \label{fig:trmline}
}\hspace{2em}
\subfigure[]{
  \includegraphics[width=.4\textwidth]{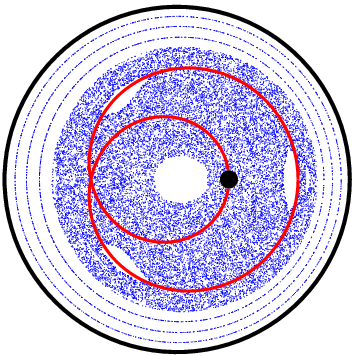}
  \label{fig:trmpoinc}
}
\caption{(a) Numerical simulation of a single rod tracing out an
  epitrochoidal path [solid line in (b)] stretching a material line
  for 7 periods.  The period-3 orbit discussed in the text is shown
  superimposed, with iterates~$\per_0$,~$\per_1$, and~$\per_2$.  (b) A
  Poincar\'e section (stroboscopic map) of some representative
  trajectories shows that the phase space consists of a large chaotic
  region and several regions of regular behaviour.}
\label{fig:trm}
\end{figure}

The protocol discussed in the present section is of the latter type.
Figure~\ref{fig:trmline} shows a material line advected by a one-rod
stirring device, where the rod follows an epitrochoidal path.
The path of the rod is shown as a solid line in
Fig.~\ref{fig:trmpoinc}, superimposed on a Poincar\'e section.  We
make two observations about Fig.~\ref{fig:trm}: (i) The Poincar\'e
section reveals a mixed phase space, consisting of a large chaotic
region and several smaller regular regions, including a regular region
that completely encloses the wall. (ii) The injection cusps into the
mixing region are not readily apparent, though small cusps are visible.

The presence of the chaotic region can be understood by examining the
motion of the
physical rod and of the regular islands visible in
Fig.~\ref{fig:trmpoinc}.  These regular islands are the ghost rods
that we use to explain the topological properties of the
homeomorphism~$\ho$ induced by the rod motion, as analyzed in
Ref.~\cite{Gouillart2006}.  The braid formed by the rod and the
ghost rods
is~$\sigma_1\sigma_2\sigma_3\sigma_2^3\sigma_3^2
\sigma_4\sigma_2\sigma_3\sigma_2^3$, which can be shown (with
software~\cite{HallTrain}, say) to correspond
to a pseudo-Anosov isotopy class with a single separatrix on the
boundary, as well as two interior 3-pronged singularities.  The
singularity data is thus~$(\bsing=1,\sing_3=2)$ for~$\nr=5$ rods ---
see Table~\ref{tab:tttypes}.

This brings us to our second observation: where is the injection cusp
associated with the boundary separatrix, as predicted by the braid?
The injection cusp is there, though it is much less evident than those
in Figs.~\ref{fig:fig8_protocol},~\ref{fig:s1s2s-3},
and~\ref{fig:morerods}.  This is because here the separatrix is
associated with a hyperbolic fixed point, as opposed to parabolic in
the previous cases.  An injection cusp near a parabolic point on the
boundary has very `slow' dynamics near the separatrix, meaning that
fluid approaches the separatrix very slowly~\cite{Gouillart2007}.
This makes the separatrices clearly visible as unmixed `tongues' in
Figs.~\ref{fig:fig8_protocol},~\ref{fig:s1s2s-3},
and~\ref{fig:morerods}.

In contrast, the injection into the mixing region is governed here by
a period-3 hyperbolic orbit near the boundary of the central mixing
region.  The orbit --- consisting of the iterates~$\per_0$, $\per_1$,
and~$\per_2$ --- is shown in Fig.~\ref{fig:trmline}.  Notice that the
orbit itself does not enter the central mixing region, just as in the
cases previously considered the parabolic fixed points at the wall
remain there.  However, a portion of the unstable manifold of each
iterate is shown
\begin{figure}
  \begin{center}
      \includegraphics[width=.45\textwidth]{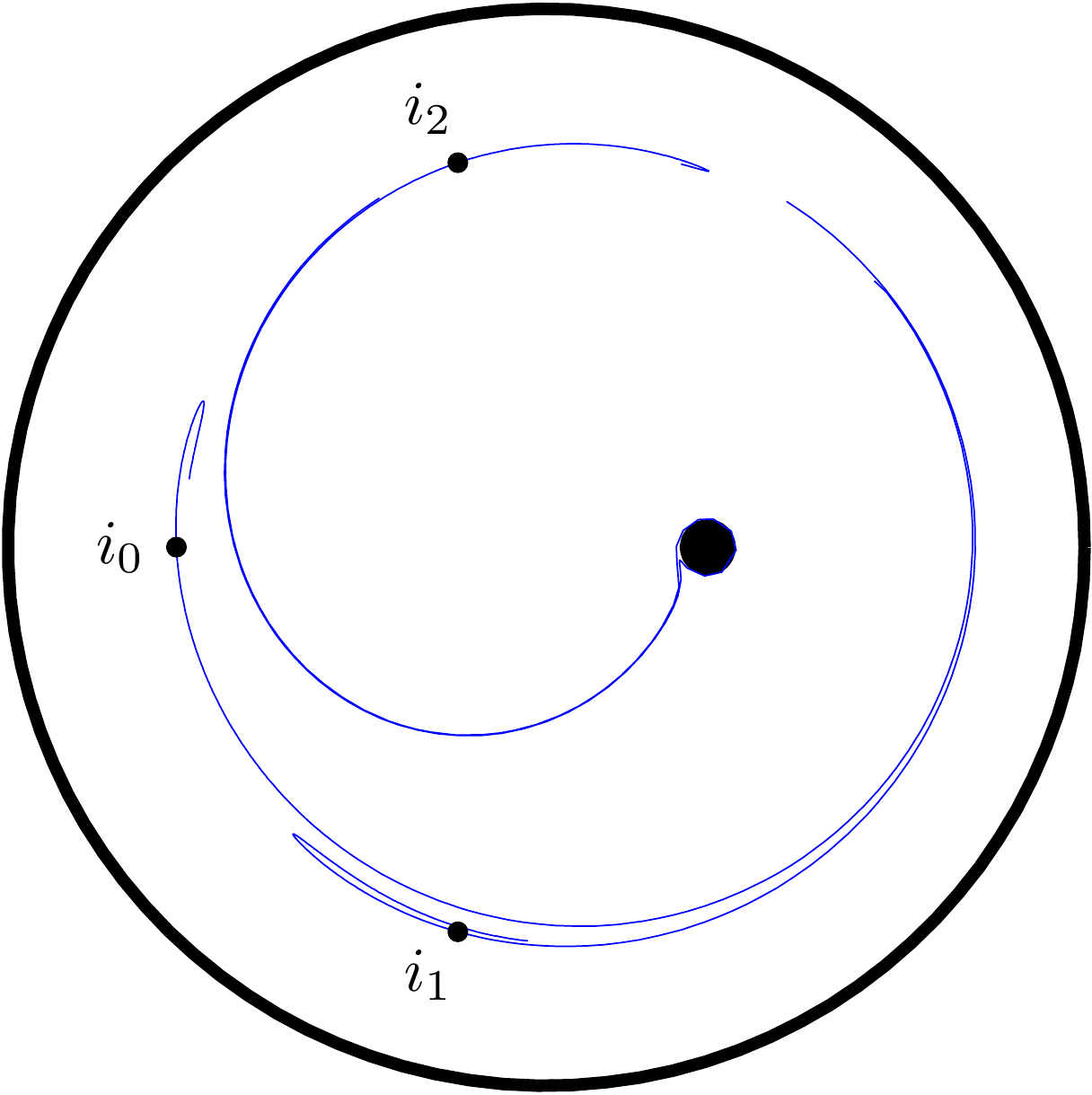}
  \end{center}
  \caption{Portions of the unstable manifold of each iterate of the
    period-3 hyperbolic orbit near the boundary of the mixing region.
  The unstable manifold of iterate~$\per_2$ enters the mixing region
  and corresponds to the location of the injection cusp.}
  \label{fig:trmp3unst}
\end{figure}%
in Fig.~\ref{fig:trmp3unst}: notice how the unstable manifold of
iterate~$\per_2$ enters the heart of the mixing region, following the
rod.  This unstable manifold is the boundary separatrix predicted by
the braid.  Thus, the way in which fluid enters the mixing region is
by coming near~$\per_2$ and then being dragged along its unstable
manifold.  The iterates~$\per_0$ and~$\per_1$ play no role as far as
injection into the mixing region is concerned.  With hindsight, we can
see the unstable manifold of~$\per_2$ in the wake of the rod in
Fig.~\ref{fig:trmline}.  Because the dynamics in their vicinity is
exponential rather than algebraic, hyperbolic injection cusps can
dramatically speed up the rate of mixing~\cite{Gouillart_thesis} in
the central region.  The price to pay is an unmixed region around the
wall of the device.

\section{Discussion}
\label{sec:discussion}

A stirring device consisting of moving rods undergoing periodic motion
induces a mapping of the fluid domain to itself.  This mapping can be
regarded as a homeomorphism of a punctured surface to itself, where
the punctures mimic the moving rods.  Having the rods undergo a
complex braiding motion guarantees a minimal amount of topological
entropy, where by complex we mean that the isotopy class associated
with the braid is pseudo-Anosov.  The topological entropy is itself a
lower bound on the rate of stretching of material lines, a quantity
which is important for chaotic mixing.

Topological considerations also predict the nature of the injection of
unmixed material into the central mixing region.  The number of
boundary separatrices in the pseudo-Anosov
homeomorphism's unstable foliation determines the number of such
injection cusps.  The number and position of injection cusps is
particularly important for open flows, such as flows in channels,
since there the nature of injection has a profound impact on the shape
of the downstream mixing pattern (Fig.~\ref{fig:channel_exp}).

Topological index formulas allow us to predict the possible types of
unstable foliations that can occur for a fixed number of rods.  We did
not provide a way of deriving the topological type of the unstable
foliation for a given rod stirring protocol.  This can be done, for
instance, by using an implementation~\cite{HallTrain} of the
Bestvina--Handel algorithm~\cite{Bestvina1995}.  Using the enumeration
presented here, a mixing device can be designed with a specific number
of injection cusps into the mixing region, by allowing for enough rods
and choosing the appropriate stirring protocol.

More generally, instead of physical rods we can consider periodic
orbits associated with a stirring protocol.  We call such periodic
orbits `ghost rods' when they play a similar role to physical rods
(that is, material lines fold around them as if they were
rods)~\cite{Thiffeault2005,Thiffeault2005preprint,%
  Gouillart2006,Stremler2007,Binder2008}.  The topological description
presented here applies to the unstable foliation associated with
periodic orbits.

In future work, we will consider not just the number of injection
cusps, but their relative position as well.  Indeed, observe that in
Fig~\ref{fig:s1s-2s3s-4s3s-2} the position of the injection cusps
alternates sides relative to the line of rods.  This is thought to be
a general feature of pseudo-Anosov stirring protocols, but the proof
of this requires careful consideration of whether or not given
foliations are \emph{dynamically} allowable, in the sense that they
can be realised as the unstable foliation of a pseudo-Anosov
homeomorphism.

\section*{Acknowledgments}

Several of the authors were first introduced to these ideas by Philip
Boyland, to whom we are extremely grateful for many stimulating
discussions.


\begin{thebibliography}{33}
\newcommand{\enquote}[1]{`#1'}
\providecommand{\natexlab}[1]{#1}
\providecommand{\url}[1]{\texttt{#1}}
\providecommand{\urlprefix}{URL }
\providecommand{\bibinfo}[2]{#2}
\providecommand{\eprint}[2][]{\url{#2}}

\bibitem[{Gouillart \emph{et~al.}(2007)Gouillart, Kuncio, Dauchot, Dubrulle,
  Roux, and Thiffeault}]{Gouillart2007}
\bibinfo{author}{E.~Gouillart}, \bibinfo{author}{N.~Kuncio},
  \bibinfo{author}{O.~Dauchot}, \bibinfo{author}{B.~Dubrulle},
  \bibinfo{author}{S.~Roux}, and \bibinfo{author}{J.-L. Thiffeault},
  \enquote{\bibinfo{title}{Walls inhibit chaotic mixing},}
  \emph{\bibinfo{journal}{Phys. Rev. Lett.}} \textbf{\bibinfo{volume}{99}},
  \bibinfo{pages}{114501} (\bibinfo{year}{2007}).

\bibitem[{Aref(1984)}]{Aref1984}
\bibinfo{author}{H.~Aref}, \enquote{\bibinfo{title}{Stirring by chaotic
  advection},} \emph{\bibinfo{journal}{J. Fluid Mech.}}
  \textbf{\bibinfo{volume}{143}}, \bibinfo{pages}{1--21}
  (\bibinfo{year}{1984}).

\bibitem[{Aref(2002)}]{Aref2002}
\bibinfo{author}{H.~Aref}, \enquote{\bibinfo{title}{The development of chaotic
  advection},} \emph{\bibinfo{journal}{Phys. Fluids}}
  \textbf{\bibinfo{volume}{14}}~(\bibinfo{number}{4}),
  \bibinfo{pages}{1315--1325} (\bibinfo{year}{2002}).

\bibitem[{Fathi \emph{et~al.}(1979)Fathi, Laundenbach, and
  Po\'{e}naru}]{Fathi1979}
\bibinfo{author}{A.~Fathi}, \bibinfo{author}{F.~Laundenbach}, and
  \bibinfo{author}{V.~Po\'{e}naru}, \enquote{\bibinfo{title}{Travaux de
  {T}hurston sur les surfaces},} \emph{\bibinfo{journal}{Ast\'{e}risque}}
  \textbf{\bibinfo{volume}{66-67}}, \bibinfo{pages}{1--284}
  (\bibinfo{year}{1979}).

\bibitem[{Thurston(1988)}]{Thurston1988}
\bibinfo{author}{W.~P. Thurston}, \enquote{\bibinfo{title}{On the geometry and
  dynamics of diffeomorphisms of surfaces},} \emph{\bibinfo{journal}{Bull. Am.
  Math. Soc.}} \textbf{\bibinfo{volume}{19}}, \bibinfo{pages}{417--431}
  (\bibinfo{year}{1988}).

\bibitem[{Boyland \emph{et~al.}(2000)Boyland, Aref, and Stremler}]{Boyland2000}
\bibinfo{author}{P.~L. Boyland}, \bibinfo{author}{H.~Aref}, and
  \bibinfo{author}{M.~A. Stremler}, \enquote{\bibinfo{title}{Topological fluid
  mechanics of stirring},} \emph{\bibinfo{journal}{J. Fluid Mech.}}
  \textbf{\bibinfo{volume}{403}}, \bibinfo{pages}{277--304}
  (\bibinfo{year}{2000}).

\bibitem[{Gouillart \emph{et~al.}(2006)Gouillart, Finn, and
  Thiffeault}]{Gouillart2006}
\bibinfo{author}{E.~Gouillart}, \bibinfo{author}{M.~D. Finn}, and
  \bibinfo{author}{J.-L. Thiffeault}, \enquote{\bibinfo{title}{Topological
  mixing with ghost rods},} \emph{\bibinfo{journal}{Phys. Rev. E}}
  \textbf{\bibinfo{volume}{73}}, \bibinfo{pages}{036311}
  (\bibinfo{year}{2006}).

\bibitem[{Boyland \emph{et~al.}(2003)Boyland, Stremler, and Aref}]{Boyland2003}
\bibinfo{author}{P.~L. Boyland}, \bibinfo{author}{M.~A. Stremler}, and
  \bibinfo{author}{H.~Aref}, \enquote{\bibinfo{title}{Topological fluid
  mechanics of point vortex motions},} \emph{\bibinfo{journal}{Physica D}}
  \textbf{\bibinfo{volume}{175}}, \bibinfo{pages}{69--95}
  (\bibinfo{year}{2003}).

\bibitem[{Vikhansky(2003)}]{Vikhansky2003}
\bibinfo{author}{A.~Vikhansky}, \enquote{\bibinfo{title}{Chaotic advection of
  finite-size bodies in a cavity flow},} \emph{\bibinfo{journal}{Phys. Fluids}}
  \textbf{\bibinfo{volume}{15}}~(\bibinfo{number}{7}),
  \bibinfo{pages}{1830--1836} (\bibinfo{year}{2003}).

\bibitem[{Finn \emph{et~al.}(2003{\natexlab{a}})Finn, Cox, and
  Byrne}]{MattFinn2003}
\bibinfo{author}{M.~D. Finn}, \bibinfo{author}{S.~M. Cox}, and
  \bibinfo{author}{H.~M. Byrne}, \enquote{\bibinfo{title}{Topological chaos in
  inviscid and viscous mixers},} \emph{\bibinfo{journal}{J. Fluid Mech.}}
  \textbf{\bibinfo{volume}{493}}, \bibinfo{pages}{345--361}
  (\bibinfo{year}{2003}{\natexlab{a}}).

\bibitem[{Finn \emph{et~al.}(2003{\natexlab{b}})Finn, Cox, and
  Byrne}]{MattFinn2003b}
\bibinfo{author}{M.~D. Finn}, \bibinfo{author}{S.~M. Cox}, and
  \bibinfo{author}{H.~M. Byrne}, \enquote{\bibinfo{title}{Chaotic advection in
  a braided pipe mixer},} \emph{\bibinfo{journal}{Phys. Fluids}}
  \textbf{\bibinfo{volume}{15}}~(\bibinfo{number}{11}),
  \bibinfo{pages}{L77--L80} (\bibinfo{year}{2003}{\natexlab{b}}).

\bibitem[{Thiffeault(2005)}]{Thiffeault2005}
\bibinfo{author}{J.-L. Thiffeault}, \enquote{\bibinfo{title}{Measuring
  topological chaos},} \emph{\bibinfo{journal}{Phys. Rev. Lett.}}
  \textbf{\bibinfo{volume}{94}}~(\bibinfo{number}{8}), \bibinfo{pages}{084502}
  (\bibinfo{year}{2005}).

\bibitem[{Thiffeault and Finn(2006)}]{Thiffeault2006}
\bibinfo{author}{J.-L. Thiffeault} and \bibinfo{author}{M.~D. Finn},
  \enquote{\bibinfo{title}{Topology, braids, and mixing in fluids},}
  \emph{\bibinfo{journal}{Phil. Trans. R. Soc. Lond. A}}
  \textbf{\bibinfo{volume}{364}}, \bibinfo{pages}{3251--3266}
  (\bibinfo{year}{2006}).

\bibitem[{Kobayashi and Umeda(2007)}]{Kobayashi2007}
\bibinfo{author}{T.~Kobayashi} and \bibinfo{author}{S.~Umeda},
  \enquote{\bibinfo{title}{Realizing pseudo-{A}nosov egg beaters with simple
  mecanisms},} in \emph{\bibinfo{booktitle}{Proceedings of the International
  Workshop on Knot Theory for Scientific Objects, Osaka, Japan}},
  \bibinfo{pages}{97--109} (\bibinfo{publisher}{Osaka Municipal Universities
  Press}, \bibinfo{year}{2007}).

\bibitem[{Binder and Cox(2008)}]{Binder2008}
\bibinfo{author}{B.~J. Binder} and \bibinfo{author}{S.~M. Cox},
  \enquote{\bibinfo{title}{A mixer design for the pigtail braid},}
  \emph{\bibinfo{journal}{Fluid Dyn. Res.}} \textbf{\bibinfo{volume}{49}},
  \bibinfo{pages}{34--44} (\bibinfo{year}{2008}).

\bibitem[{Finn \emph{et~al.}(2006)Finn, Thiffeault, and
  Gouillart}]{MattFinn2006}
\bibinfo{author}{M.~D. Finn}, \bibinfo{author}{J.-L. Thiffeault}, and
  \bibinfo{author}{E.~Gouillart}, \enquote{\bibinfo{title}{Topological chaos in
  spatially periodic mixers},} \emph{\bibinfo{journal}{Physica D}}
  \textbf{\bibinfo{volume}{221}}~(\bibinfo{number}{1}),
  \bibinfo{pages}{92--100} (\bibinfo{year}{2006}).

\bibitem[{Finn and Thiffeault(2007{\natexlab{a}})}]{MattFinn2007}
\bibinfo{author}{M.~D. Finn} and \bibinfo{author}{J.-L. Thiffeault},
  \enquote{\bibinfo{title}{Topological entropy of braids on the torus},}
  \emph{\bibinfo{journal}{SIAM J. Appl. Dyn. Syst.}}
  \textbf{\bibinfo{volume}{6}}, \bibinfo{pages}{79--98}
  (\bibinfo{year}{2007}{\natexlab{a}}).

\bibitem[{Jana \emph{et~al.}(1994)Jana, Metcalfe, and Ottino}]{Jana1994b}
\bibinfo{author}{S.~C. Jana}, \bibinfo{author}{G.~Metcalfe}, and
  \bibinfo{author}{J.~M. Ottino}, \enquote{\bibinfo{title}{Experimental and
  computational studies of mixing in complex {S}tokes flows: the vortex mixing
  flow and multicellular cavity flow},} \emph{\bibinfo{journal}{J. Fluid
  Mech.}} \textbf{\bibinfo{volume}{269}}, \bibinfo{pages}{199--246}
  (\bibinfo{year}{1994}).

\bibitem[{Birman(1975)}]{Birman1975}
\bibinfo{author}{J.~S. Birman}, \emph{\bibinfo{title}{Braids, Links, and
  Mapping Class Groups}}, Annals of Mathematics Studies
  (\bibinfo{publisher}{Princeton University Press},
  \bibinfo{address}{Princeton, NJ}, \bibinfo{year}{1975}).

\bibitem[{Handel(1985)}]{Handel1985}
\bibinfo{author}{M.~Handel}, \enquote{\bibinfo{title}{Global shadowing of
  pseudo-{A}nosov homeomorphisms},} \emph{\bibinfo{journal}{Ergod. Th. Dynam.
  Syst.}} \textbf{\bibinfo{volume}{8}}, \bibinfo{pages}{373--377}
  (\bibinfo{year}{1985}).

\bibitem[{Boyland(1999)}]{Boyland1999}
\bibinfo{author}{P.~L. Boyland}, \enquote{\bibinfo{title}{Isotopy stability of
  dynamics on surfaces},} in \emph{\bibinfo{booktitle}{Geometry and Topology in
  Dynamics (Winston-Salem, {NC}, 1998/San Antonio, {TX}, 1999)}},
  \emph{\bibinfo{series}{Contemp. Math.}}, volume \bibinfo{volume}{246},
  \bibinfo{pages}{17--45} (\bibinfo{publisher}{American Mathematical Society},
  \bibinfo{address}{Providence, {RI}}, \bibinfo{year}{1999}).

\bibitem[{Arnold and Avez(1968)}]{ArnoldAvez}
\bibinfo{author}{V.~I. Arnold} and \bibinfo{author}{A.~Avez},
  \emph{\bibinfo{title}{Ergodic Problems of Classical Mechanics}}
  (\bibinfo{publisher}{W. A. Benjamin}, \bibinfo{address}{New York},
  \bibinfo{year}{1968}).

\bibitem[{Penner and Harer(1991)}]{PennerHarer}
\bibinfo{author}{R.~C. Penner} and \bibinfo{author}{J.~L. Harer},
  \emph{\bibinfo{title}{Combinatorics of Train Tracks}}, number
  \bibinfo{number}{125} in \bibinfo{series}{Annals of Mathematics Studies}
  (\bibinfo{publisher}{Princeton University Press},
  \bibinfo{address}{Princeton, NJ}, \bibinfo{year}{1991}).

\bibitem[{Band and Boyland(2007)}]{BandBoyland2007}
\bibinfo{author}{G.~Band} and \bibinfo{author}{P.~L. Boyland},
  \enquote{\bibinfo{title}{The {B}urau estimate for the entropy of a braid},}
  \emph{\bibinfo{journal}{Algeb. Geom. Topology}} \textbf{\bibinfo{volume}{7}},
  \bibinfo{pages}{1345--1378} (\bibinfo{year}{2007}).

\bibitem[{Milnor(1997)}]{Milnor}
\bibinfo{author}{J.~W. Milnor}, \emph{\bibinfo{title}{Topology from the
  Differentiable Viewpoint}}, \bibinfo{edition}{revised} edition
  (\bibinfo{publisher}{Princeton University Press},
  \bibinfo{address}{Princeton, NJ}, \bibinfo{year}{1997}).

\bibitem[{Thurston(1997)}]{Thurston3D1}
\bibinfo{author}{W.~P. Thurston}, \emph{\bibinfo{title}{Three-dimensional
  geometry and topology}}, volume~\bibinfo{volume}{1}
  (\bibinfo{publisher}{Princeton University Press},
  \bibinfo{address}{Princeton, NJ}, \bibinfo{year}{1997}),
  \bibinfo{note}{edited by Silvio Levy}.

\bibitem[{Andrews(1976)}]{AndrewsPartitions}
\bibinfo{author}{G.~E. Andrews}, \emph{\bibinfo{title}{The Theory of
  Partitions}} (\bibinfo{publisher}{Addison-Wesley}, \bibinfo{address}{Reading,
  Mass.}, \bibinfo{year}{1976}).

\bibitem[{Stremler and Chen(2007)}]{Stremler2007}
\bibinfo{author}{M.~A. Stremler} and \bibinfo{author}{J.~Chen},
  \enquote{\bibinfo{title}{Generating topological chaos in lid-driven cavity
  flow},} \emph{\bibinfo{journal}{Phys. Fluids}} \textbf{\bibinfo{volume}{19}},
  \bibinfo{pages}{103602} (\bibinfo{year}{2007}).

\bibitem[{Hall(0000)}]{HallTrain}
\bibinfo{author}{T.~Hall}, \enquote{\bibinfo{title}{\textit{Train: {A} {C++}
  program for computing train tracks of surface homeomorphisms}},}
  \bibinfo{note}{\texttt{http://www.liv.ac.uk/maths/PURE/MIN\_SET/CONTENT/memb%
ers/T\_Hall.html}}.

\bibitem[{Gouillart(2007)}]{Gouillart_thesis}
\bibinfo{author}{E.~Gouillart}, \emph{\bibinfo{title}{Chaotic mixing by
  rod-stirring devices in open and closed flows}}, Ph.D. thesis,
  \bibinfo{school}{Universit\'{e} Pierre et Marie Curie Paris 6}
  (\bibinfo{year}{2007}),
  \eprint{http://tel.archives-ouvertes.fr/tel-00204109/en}.

\bibitem[{Bestvina and Handel(1995)}]{Bestvina1995}
\bibinfo{author}{M.~Bestvina} and \bibinfo{author}{M.~Handel},
  \enquote{\bibinfo{title}{Train-tracks for surface homeomorphisms},}
  \emph{\bibinfo{journal}{Topology}}
  \textbf{\bibinfo{volume}{34}}~(\bibinfo{number}{1}),
  \bibinfo{pages}{109--140} (\bibinfo{year}{1995}).

\bibitem[{Thiffeault \emph{et~al.}(2005)Thiffeault, Gouillart, and
  Finn}]{Thiffeault2005preprint}
\bibinfo{author}{J.-L. Thiffeault}, \bibinfo{author}{E.~Gouillart}, and
  \bibinfo{author}{M.~D. Finn}, \enquote{\bibinfo{title}{The size of ghost
  rods},}  (\bibinfo{year}{2005}), \eprint{arXiv:nlin/0507076}.

\end{thebibliography}

\end{document}